\documentclass[twocolumn,aps,prc,tightenlines,floats,floatfix,eqsecnum,preprintnumbers,nofootinbib]{revtex4}

\def\sp{\kern +3pt}
\def\sm{\kern -3pt}
\def\spQ{\kern +6pt}

\def\bea{\begin{eqnarray}}
\def\eea{\end{eqnarray}}

\def\sfrac#1#2{{\textstyle \frac{#1}{#2}}}

\newcommand{\bra}[1]{\langle #1|}
\newcommand{\ket}[1]{|#1\rangle}

\newcommand{\unit}{1\!\!1}

\def\be{\begin{equation}}
\def\ee{\end{equation}}
\def\ba{\begin{eqnarray}}
\def\ea{\end{eqnarray}}

\usepackage{graphics}
\usepackage{graphicx}
\usepackage{epsf}
\usepackage{amsmath}
\usepackage{amssymb}

\begin{document}

\phantom{0}
\vspace{-0.2in}
\hspace{5.5in}

\preprint{ }

\vspace{-1in}

\title
{\bf $\gamma^\ast N \to N(1710)$ transition
at high momentum transfer}
\author{G.~Ramalho and
K.~Tsushima}
\vspace{-0.1in}

\affiliation{
International Institute of Physics, Federal
University of Rio Grande do Norte, Avenida Odilon Gomes de Lima 1722,
Capim Macio, Natal-RN 59078-400, Brazil
}

\vspace{0.2in}
\date{\today}

\phantom{0}

\begin{abstract}
In a relativistic quark model
we study the structure of the $N(1710)$ resonance,
and the $\gamma^\ast N \to N(1710)$ reaction
focusing on the high momentum transfer region, where
the valence quark degrees of freedom are expected to be dominant.
The $N(1710)$ resonance, a state with spin 1/2
and positive parity  ($J^P = \frac{1}{2}^+$),
can possibly be interpreted as the second radial excitation of
the nucleon, after the Roper, $N(1440)$.
We calculate the $\gamma^\ast N \to N(1710)$ helicity amplitudes,
and predict that they are almost identical to those of the
$\gamma^\ast N \to N(1440)$ reaction in the high momentum transfer region.
Thus, future measurement of the helicity amplitudes for
the $\gamma^\ast N \to N(1710)$ reaction
can give a significant hint on the internal structure
of the $N(1710)$ state.
\end{abstract}

\vspace*{0.9in}  
\maketitle

\section{Introduction}

The understanding of the electromagnetic structure
of the hadrons, and its connection
with the underlying degrees of freedom in Quantum Chromodynamics (QCD),
are amongst, two of the more interesting challenges
in the hadronic physics.
Through the nucleon electro-excitation reactions,
$e N \to e' N^\ast$, we can study the electromagnetic structure of
the nucleon ($N$) and the nucleon excitations ($N^\ast$),
including also the $\Delta$ states.
The excitation of nucleon resonances occurs through
the intermediate processes $\gamma^\ast N \to N^\ast$,
where $\gamma^\ast$ is a virtual photon,
and the cross sections of the processes can usually be
expressed in terms of the electromagnetic transition form factors,
or helicity transition amplitudes.
Those functions depend on the four-momentum transfer
squared $q^2$, and characterize the electromagnetic
structure of such resonant states.

The data extracted from experiments
in facilities such as Jefferson Lab (JLab) and MAMI (Mainz),
allow us to extract the electromagnetic
transition form factors of resonant states
in the region $Q^2 < 5$ GeV$^2$ ($Q^2=-q^2$),
corresponding to the first and second
resonance region $W  <  1.6$ GeV, with
$W$ being the $\gamma^\ast N$ invariant mass~\cite{Aznauryan12a,NSTAR}.
The planned JLab 12-GeV upgrade
will enable us to make a detailed study of
the resonances in the third-resonance region
($W \approx 1.7$ GeV)~\cite{NSTAR,Jlab12GeV}.

There is some evidence for the existence of a state $N(1710)$
with quantum numbers $J^P=\frac{1}{2}^+$ since
the eighties~\cite{Cutkosky79,Hohler82}.
(In the notation of the $\pi N$ scattering
it is labeled as $N(1710) P_{11}$~\cite{PDG}.)
The state $N(1710)  \frac{1}{2}^+$ is now classified as a three-star resonance,
and found by several groups~\cite{Anisovich12,Shrestha12,Ceci06a,Ceci06b,Osmanovic11,Shklyar07,MAID02,Batinic95,Feuster98}
in their partial wave analyses,
and it is also included in some baryon-meson reaction models~\cite{EBAC,Kamano13,Ronchen13,Yang12,Vrana00a,Chiang04,Khemchandani08}.
The resonance identified as
$N(1710)  \frac{1}{2}^+$ has a small decay branching ratio
for the $\pi N$ channel,
but significant decay branching ratios for the $\eta N$, $K \Lambda$
and $\pi \pi N$ channels~\cite{PDG},
and thus, may have important roles
in kaon~\cite{Kaon}
and hypernuclear production~\cite{Shyam2}.
However, in some partial wave analysis
like SAID (GWU), the resonance is not present~\cite{Arndt06}.
Thus, the clear existence of the $N(1710)$ state is still controversial,
although some authors defend that
it should be classified as a four-star resonance~\cite{Ceci06a}.
A state $N \frac{1}{2}^+$ with a mass near 1.7 GeV
collected also some interest
as a possible partner of the $\Theta^+$-pentaquark,
non-strange anti-decuplet state~\cite{Pentaquarks,Hicks12}.
However, recent experimental evidence on the existence of the
$\Theta^+$-pentaquark seems to have been weakened~\cite{Hicks12}.
Lattice QCD simulations
predict also $N\frac{1}{2}^+$ states that can
possibly be related to the $N(1710)$ state~\cite{Edwards11}.
Future experiments,
like the ones planned at the JLab-12 GeV upgraded
facility, will help to establish or deny the existence
of the $N(1710)$ state.
If the existence of $N(1710)$ state is confirmed,
measurement of the $\gamma^\ast N \to N(1710)$
helicity amplitudes will be possible.
Thus, it is  important to
present predictions at this stage for the reaction
associated with $N(1710)$, as we do in the present study.

In the usual nonrelativistic quark models
the resonances $N(1440)$ (also called Roper)
and $N(1710)$
can be classified as $N \frac{1}{2}^+$ states
with two different radial excitation of the nucleon,
although the masses are different from the
corresponding resonant poles~\cite{Capstick00,Koniuk80}.
The Roper has intriguing properties that
are difficult to explain in a context of
the quark model framework~\cite{Aznauryan12a,NSTAR,Capstick00,Koniuk80,EBAC1,Suzuki10,Wilson12}.
In a nonrelativistic quark model description
the mass of the Roper is too heavy
and the decay width is too narrow to be compatible
with a three-quark ($qqq$) system.
Therefore, several alternative descriptions
were suggested, e.g., by identifying the state
as a quark bare core combined with gluon states ($qqqG$),
a molecular-type state ($N\sigma$),
or a dynamically generated state by baryon and meson.
However, recent data from CLAS and MAID~\cite{CLAS,MAID}
for $Q^2 > 2$ GeV$^2$
support the assumption that the $N (1440)$
is predominantly the first radial excitation
of the nucleon~\cite{Aznauryan12a}.
Estimates made by constituent and light-front quark models
gave a good description for the high $Q^2$ region data.
For a review concerning the $N(1440)$ resonance,
see Ref.~\cite{Aznauryan12a}.

Based on the knowledge of the Roper,
one can raise the following question,
whether or not the $N(1710)$ state can be described
as an another radial excitation of the nucleon.
Some other works describe the state
as a three-quark
system~\cite{Capstick93,Capstick95,Capstick99,Melde08,Santopinto12,Ronniger13},
as a baryon bare core with a meson cloud~\cite{Golli08},
or $\sigma$-meson and glueball excitations~\cite{Alberto01}.
There are, however, suggestions
that the state $N(1710)$ may be more likely generated dynamically
by baryon-meson states~\cite{EBAC1,Suzuki10,Khemchandani08,Zhang12}.
The state $N(1710)$ is also predicted
in algebraic models of QCD~\cite{Bijker,Bijker2}
and in the large $N_c$ limit~\cite{Matagne05}.
The wave function of the $N(1710)$
is also estimated by lattice QCD simulations~\cite{Roberts13}.

As already recognized for the Roper
and several other resonance cases,
the meson cloud dressing for the transition
helicity amplitudes is expected to
be important only in the low $Q^2$ region~\cite{Aznauryan12a,NSTAR}.
At high $Q^2$ the valence quark degrees of freedom
are expected to  dominate
and the helicity amplitudes
follow the power laws,
$A_{1/2} \propto 1/Q^3$ and $S_{1/2} \propto 1/Q^5$~\cite{Carlson}.
In this work, focusing on the high $Q^2$ region,
we predict the form factors and the helicity amplitudes
of the $\gamma^\ast N \to N(1710)$ reaction,
using a relativistic constituent quark model
and the assumption that the $N(1710)$ state
is the second radial excitation of the nucleon state.
That is our working hypothesis.
In the near future our predictions
can thus be indeed tested in the JLab 12-GeV upgraded facility.

In the previous works the nucleon and $N(1440)$
structures were described in the covariant spectator quark
model~\cite{Nucleon,Roper} as analogous states
with different radial excitation.
In there the $N(1440)$ wave function
was determined uniquely from the nucleon wave function
by imposing
that the $N(1440)$ radial wave function
is orthogonal to the nucleon radial wave function,
without any new adjustable parameters in the model~\cite{Roper}.
In the present work we extend the procedure
to the second radial excitation of the nucleon
(first radial excitation of the Roper), $N(1710)$.
In a similar manner, the radial wave function of $N(1710)$
can be determined by the two orthogonality conditions
between the $N(1710)$ radial wave function
and both of the Roper and nucleon wave functions,
again without any new adjustable parameters.
The price we must pay for the present prediction
is that the estimates of the electromagnetic properties
are expected to be valid only in the high $Q^2$ region,
since the calculations are
based exclusively on the valence quark degrees of
freedom.

We will conclude that the transition helicity amplitudes
for the $\gamma^\ast N \to N(1710)$ reaction
have their magnitude and falloff very similar to
those for the $\gamma^\ast N \to N(1440)$ reaction for
$Q^2 > 4$ GeV$^2$.

This article is organized as follows.
In Sec.~\ref{secEMFF} we define
the transition form factors and
the helicity amplitudes.
We describe in Sec.~\ref{secCSQM}
the covariant spectator quark model,
and present the baryon wave functions and
the analytic expressions for the transition form factors and
the helicity amplitudes.
The numerical results are presented in Sec.~\ref{secResults},
and finally, the summary and the conclusions are
given in Sec.~\ref{secConclusions}.

\section{Electromagnetic form factors and helicity amplitudes}
\label{secEMFF}

The electromagnetic transition between a nucleon
(mass $M$) and a resonance $N^\ast$ (mass $M_R$) with $J^P=\frac{1}{2}^+$,
can be described by the current
(in units of the proton charge $e$)~\cite{Aznauryan12a,Roper}:
\ba
& &
\hspace{-0.5cm}
J^\mu=
\left( \gamma^\mu - \frac{\not \! q q^\mu}{q^2}\right)F_1^\ast (Q^2)
+
\frac{i \sigma^{\mu \nu} q_\nu}{M_R+ M} F_2^\ast (Q^2),
\nonumber \\
& &
\label{eqJgen}
\ea
which defines the Dirac ($F_1^\ast$) and Pauli ($F_2^\ast$)
form factors.
The current operator $J^\mu$ can be projected
on the Dirac spinors of the resonance $u_R(P_+)$ and
of the nucleon $u(P_-)$, where
$P_+$ ($P_-$) is the final (initial) momentum,
$q=P_+-P_-$ and $Q^2=-q^2$.
Spin projection indices are suppressed in the spinors
for simplicity.

The experimental data measured for hadron electromagnetic
reactions are usually reported in terms of
the helicity amplitudes in the final state ($N^\ast$) rest frame.
In this case
the current (\ref{eqJgen}) is projected
on the initial and final spin states
using the photon polarization states,
$\varepsilon_\lambda^\mu$,
with $\lambda=0,\pm$ being the photon spin projection.
In the $N^\ast$ rest frame,
one has the helicity amplitudes, $A_{1/2}$ and
$S_{1/2}$, which are given by~\cite{Aznauryan12a,Roper}:
\ba
& &
A_{1/2}(Q^2)= {\cal K}
\bra{N^\ast,+\sfrac{1}{2}} \varepsilon_+ \cdot J
\ket{N, -\sfrac{1}{2}},
\label{eqA1}\\
& &
S_{1/2}(Q^2)= {\cal K}
\bra{N^\ast,+\sfrac{1}{2}} \varepsilon_0 \cdot J
\ket{N, + \sfrac{1}{2}} \frac{|{\bf q}|}{Q},
\label{eqS0}
\ea
where
\be
{\cal K}= \sqrt{\frac{2\pi \alpha}{K}},
\ee
with $\alpha = \sfrac{e^2}{4\pi} \simeq \frac{1}{137}$,
 $K=\frac{M_R^2-M^2}{2M_R}$.
 $|{\bf q}|$ is
the photon momentum in the $N^\ast$ rest frame,
\be
|{\bf q}|= \frac{\sqrt{Q_+^2Q_-^2}}{2M_R},
\label{eqq2}
\ee
where $Q_\pm^2= (M_R \pm M)^2 + Q^2$. 

The helicity amplitudes, $A_{1/2}$ and $S_{1/2}$,
can be related with the form factors $F_1^\ast$
and $F_2^\ast$ via
Eqs.~(\ref{eqJgen})-(\ref{eqS0})~\cite{Aznauryan12a,Roper} as
\ba
& &
\hspace{-1.2cm}
A_{1/2}(Q^2)=
{\cal R} \left\{ F_1^\ast (Q^2) + F_2^\ast(Q^2) \right\},
\label{eqA12}\\
& &
\hspace{-1.2cm}
S_{1/2}(Q^2)=
\frac{ {\cal R} }{\sqrt{2}}
|{\bf q}| \frac{M_R+M}{Q^2}
\left\{ F_1^\ast (Q^2) -\tau F_2^\ast(Q^2) \right\},
\label{eqS12}
\ea
where $\tau=\sfrac{Q^2}{(M_R+M)^2}$, and
\ba
{\cal R}= \frac{e}{2}\sqrt{\frac{Q_-^2}{M_R M K}}.
\ea
Note that the amplitude $S_{1/2}$ is
determined by the virtual photons and not specified
at $Q^2=0$.
The analytic properties of the current
(\ref{eqJgen}) implies that $F_1^\ast(0)=0$~\cite{Roper}.

\section{Covariant spectator quark model}
\label{secCSQM}

The covariant spectator quark model
is derived from the formalism of
the covariant spectator theory~\cite{Gross}.
In the model a baryon $B$ is described as a
three-constituent-quark system, where one quark is free to interact
with the electromagnetic fields and the other quarks are on-mass-shell.
Integrating over the on-mass-shell momenta,
one can represent the quark pair as an on-mass-shell
diquark with effective mass $m_D$,
and the baryon as a
quark-diquark system~\cite{NSTAR,Nucleon,Omega,Nucleon2}.
The structure of the baryon is then described by a
transition vertex between the three-quark bound state
and a quark-diquark state, that describes
effectively the confinement~\cite{Nucleon,Omega}.

The baryon wave function $\Psi_B(P,k)$ is then
derived from the transition vertex
as a function of the baryon momentum $P$ and
the diquark momentum $k$,
taking into account the properties of the baryon $B$,
such as the spin and flavor.
Instead of solving dynamical equations 
to get wave functions, 
the wave functions $\Psi_B$
are built from the baryon internal symmetries,
with the shape determined directly by
experimental or lattice data for 
some ground state systems~\cite{NSTAR,Nucleon,NDelta,LatticeD}.
The baryon mass $M_B$ is a parameter fixed by the experimental value.
In particular the parametrization of the nucleon wave 
function was calibrated by the nucleon electromagnetic 
form factor data~\cite{Nucleon}.

In the past the covariant spectator
quark model was applied to several nucleon
resonances such as $N(1520)$~\cite{D13}, $N(1535)$~\cite{S11},
$\Delta$ resonances
\cite{NDelta,LatticeD,Delta,Lattice,Delta1600}, and also to
other reactions~\cite{Omega,OctetFF,Octet2Decuplet,Omega2}.

\subsection{Transition current}

Once the baryon wave functions are written
in terms of the single quark and
quark-pair states, one can write
the transition current
between the baryons $B$ and $B'$
in a relativistic impulse approximation as~\cite{Nucleon,Omega,Nucleon2}
\ba
J^\mu_{B'B} =
3 \sum_{\Gamma} \int_k \bar \Psi_{B'} (P_+,k)
j_q^\mu \Psi_{B} (P_-,k),
\label{eqJ1}
\ea
where $j_q^\mu$ is the (single) quark current operator,
$P_-$ ($P_+$) is the initial (final) momentum
and $\Gamma$ labels the scalar diquark
and vectorial diquark (projections $\Lambda=0,\pm$)
polarizations.
The factor 3 takes account of the contributions
from the other quark-pairs by the symmetry,
and the integration symbol represents the
covariant integration for the diquark on-shell state
$\int_k \equiv \int \frac{d^3 {\bf k}}{(2\pi)^3(2E_D)}$,
with $E_D=\sqrt{m_D^2 + {\bf k}^2}$.
Compared to Eq.~(\ref{eqJgen}), the current $J^\mu$
is in this case projected on the initial
and final states, and labeled respectively by the indices $B$ and $B'$.

The quark current operator is expressed in
terms of the Dirac ($j_1$) and Pauli ($j_2$) quark
form factors~\cite{Nucleon,Omega}:
\ba
j_q^\mu = j_1 (Q^2) \left(  \gamma^\mu - \frac{\not \! q q^\mu}{q^2}
\right) +
j_2 (Q^2) \frac{i \sigma^{\mu \nu} q_\nu}{2M}.
\ea
The inclusion of the terms $- \frac{\not q q^\mu}{q^2}$
associated with the Dirac component
is equivalent with using the Landau prescription
for the current $J^\mu_{B'B}$~\cite{Kelly98,Batiz98}.
This terms restores current conservation,
but does not affect the results for the observables~\cite{Kelly98}.
For the case where the baryons are composed
only of $u$ and $d$ quarks,
the quark form factors $j_i$ ($i=1,2$)
can be decomposed into an isoscalar and an isovector components,
given respectively by the functions $f_{i+}$ and $f_{i-}$:
\ba
j_i= \frac{1}{6} f_{i+} (Q^2)+ \frac{1}{6} f_{i-} (Q^2)\tau_3.
\label{eqJi}
\ea
The quark form factors
are parameterized based on the
vector meson dominance mechanism~\cite{Nucleon,Omega,Lattice,OctetFF}.
The functions $f_{i\pm}$ are constrained at $Q^2=0$
in order to reproduce the quark charges and so as to
properly parameterize the anomalous magnetic moments of
the constituent quarks.
The details are given in Appendix~\ref{appSQM}.

\subsection{Baryon wave functions}

Next, we discuss the wave functions of
the nucleon and the nucleon radial excited states,
$N(1440)$ and $N(1710)$,
which will respectively be denoted by $N0, N1$ and $N2$.
In the covariant spectator quark model
the nucleon and its radial excitation
can be described as a quark-diquark system in an
$S$-state configuration~\cite{Nucleon}.
Following Refs.~\cite{Nucleon,Roper},
we can represent the $Nj$ ($j=0,1,2$)
wave functions as
\ba
\Psi_{Nj}(P,k)=
\frac{1}{\sqrt{2}}
\left[ \phi_I^0 \phi_S^0 +  \phi_I^1 \phi_S^1
\right] \psi_{Nj}(P,k),
\label{eqPsiNj}
\ea
where $\phi^{0,1}_{S}$ and  $\phi^{0,1}_{I}$
represent respectively the spin ($S$) and
isospin ($I$) states corresponding to the
total magnitude of either 0 or 1 in the diquark configuration~\cite{Nucleon}.
(See Appendix~\ref{appSQM} for detail.)
Again, we suppress the spin
and isospin projection indices
for simplicity.
The wave function represented by Eq.~(\ref{eqPsiNj})
satisfies the Dirac equation~\cite{Nucleon,NDelta}.
The functions $\psi_{Nj}(P,k)$ are the radial wave functions
to be described next.

The radial wave functions $\psi_{Nj}$
depend on the angular momentum
and the radial excitation of the baryons.
Since the baryon and the diquark are
both on-mass-shell, one can write
the radial wave functions $\psi_{Nj}$
for the quark-diquark system as
a function of $(P-k)^2$
(or $P \cdot k$)~\cite{Nucleon}.
To take into account the dependence
on a generic baryon $B$, or its mass $M_B$,
one can use the dimensionless variable:
\ba
\chi_{B} = \frac{(M_B-m_D)^2-(P- k)^2}{M_B m_D}.
\ea
In the present study $B=N0,N1$ or $N2$.
For the nucleon mass ($M_{N0}$) we also use
$M$ for simplicity.

For the nucleon radial wave function, we use~\cite{Nucleon}
\ba
\psi_{N0}(\chi_N)=
N_0\frac{1}{m_D(\beta_1 + \chi_N)(\beta_2 + \chi_N)},
\label{eqPsiN0}
\ea
where $N_0$ is the normalization constant, and
the parameter values are $\beta_1= 0.049$ and $\beta_2= 0.717$,
which were determined by the fit to the nucleon
electromagnetic form factors~\cite{Nucleon}.
If we choose the momentum-range parameters such that
$\beta_1 < \beta_2$, $\beta_1$ regulates
the spacial long-range structure.

For the Roper ($N1$), we take the form~\cite{Roper}
\ba
\psi_{N1}(\chi_{N1})=
N_1 \frac{\beta_3 - \chi_{N1}}{\beta_1 + \chi_{N1}}
\frac{1}{m_D(\beta_1 + \chi_{N1})(\beta_2 + \chi_{N1})},
\nonumber \\
\label{eqPsiN1}
\ea
where $N_1$ is the normalization constant,
and $\beta_3$ is a new parameter, fixed
by the orthogonality condition with the nucleon state.
As explained in Ref.~\cite{Roper},
the term $(\beta_3 - \chi_{N1})$ represents the
radial excitation in the momentum space
(term on $(P-k)^2$ or $\chi_{N1}$).

Finally for the $N(1710)$ ($N2$), we define
the radial wave function as
\ba
\psi_{N2}(\chi_{N2}) &=&
N_2 \frac{\chi_{N2}^2 - \beta_4 \chi_{N2} + \beta_5}{(\beta_1 + \chi_{N2})^2}
 \nonumber \\
& & \times
\frac{1}{m_D(\beta_1 + \chi_{N2})(\beta_2 + \chi_{N2})},
\label{eqPsiN2}
\ea
where $N_2$ is the normalization constant.
In this case we have two additional parameters
$\beta_4$ and $\beta_5$ to be fixed.
The minus sign in the coefficient $\beta_4$
is introduced by convenience.

In our model the sign of the normalization
constants $N_l$ ($l=0,1,2$) cannot be predicted.
In the previous works~\cite{Nucleon,Roper}
the relative sign of $N_0$ and $N_1$ was determined
by the sign of the form factors.
In the present case since there are no
available data for the high $Q^2$ region, we have
no way of fixing the sign of $N_2$ based on an experimental basis.
We assume here that $N_2$ is positive.
If future experimental data
reveal an opposite sign,
we should correct the signs of
the corresponding amplitudes.

The normalization of the radial wave functions
$\psi_{Nj}$ is determined by the
normalization of the wave function (\ref{eqPsiNj})
in order to obtain the corresponding charge of the state.
The explicit expression is
\ba
& &
\sum_{\Gamma} \int_k \bar \Psi_{Nj}(\bar P,k) j_1 \gamma^0
\Psi_{Nj}(\bar P,k) \nonumber \\
& & = \sfrac{1}{2}(1+ \tau_3)
\int_k |\psi_{Nj}(\bar P,k)|^2.
\ea
In the above
$\bar P$ is the baryon momentum at its
rest frame, $j_1 = \frac{1}{6} + \frac{1}{2} \tau_3$
is the quark charge operator in the $Q^2=0$ limit.
In order to obtain the baryon charge $\sfrac{1}{2}(1+ \tau_3)$
correctly, we need the normalization condition,
\ba
\int_k |\psi_{Nj}(\bar P,k)|^2 =1.
\ea

The orthogonality among the $N0$, $N1$ and $N2$ states
is derived from
\ba
\sum_{\Gamma} \int_k \bar \Psi_{Nj'}(\bar P_+,k) j_1 \gamma^0
\Psi_{Nj}(\bar P_-,k)
= 0,
\label{eqOrthCond1}
\ea
where $\bar P_+$ and $\bar P_-$ are
the momenta of $Nj'$ and $Nj$ respectively,
for $Q^2=0$.

\subsection{Valence quark contributions
for the electromagnetic form factors}
\label{secBareFF}

In order to write the expressions
for the form factors,
it is convenient to project the
quark current $j_i$ on the
isospin states using
$j_i^A = (\phi_I^0)^\dagger j_i \phi_I^0$, and
$j_i^S = (\phi_I^1)^\dagger j_i \phi_I^1$:
\ba
j_i^A &=& \frac{1}{6} f_{i+} +  \frac{1}{2} f_{i-} \tau_3,
\label{eqJA}
\\
j_i^S &=& \frac{1}{6} f_{i+} -  \frac{1}{6} f_{i-} \tau_3.
\ea
The results for the form factors are then given by
\ba
& &
F_{1}^\ast(Q^2) = \nonumber \\
& &
\left[\frac{3}{2} j_1^A +
\frac{1}{2}
\frac{3 -\tau}{1+ \tau}
j_1^S - 2 \frac{\tau}{1+\tau} \frac{M_{N2}+ M}{2M}
j_2^S \right] {\cal I}(Q^2),
\nonumber \\
\label{eqF1b} \\
& &F_2^\ast(Q^2) = \nonumber \\
& &
\left[
\left(
\frac{3}{2}
j_2^A
-\frac{1}{2} \frac{1-3\tau}{1+\tau} j_2^S \right) \frac{M_{N2}+ M}{2M}
-2 \frac{1}{1+\tau} j_1^S
\right] {\cal I}  (Q^2) , \nonumber \\
& &
\label{eqF2b}
\ea
with $\tau=\sfrac{Q^2}{(M_{N2} + M)^2}$, and
\be
{\cal I}(Q^2)= \int_k \psi_{N2}(P_+,k) \psi_{N0}(P_-,k),
\label{eqI20}
\ee
is the overlap integral between
the initial and final radial wave functions.
The integral ${\cal I}$ is frame independent
and is discussed in next section.

The Eqs.~(\ref{eqF1b}) and (\ref{eqF2b})
are equivalent to the
ones presented in the study of
$\gamma^\ast N \to N (1440)$~\cite{Roper},
although they are shown in different forms.
The analytic expressions obtained for
$F_1^\ast$ and $F_2^\ast$ are consistent
with the results obtained for other systems
with the same spin structure~\cite{Nucleon,OctetFF}.
If we replace $\psi_{N2} \to \psi_N$
and $M_{N2} + M \to 2M$,
we recover the expressions
for the nucleon elastic form factors~\cite{Nucleon}.
The expressions can also be related
with the octet baryon elastic form factors.
Except for the fact that $j_i^A$ and $j_i^S$
can depend also on the strange quark,
the expressions are the same when
replacing $\psi_{N2}, \psi_N \to \psi_B$
and $M_{N2} + M \to 2 M_B$,
with $\psi_B$ and $M_B$ being the octet baryon
radial wave functions and their masses respectively~\cite{OctetFF}.

\subsection{Orthogonality between the states}

The consequence of the condition~(\ref{eqOrthCond1})
is that the orthogonality between the states
is replaced by the condition for the
radial wave functions~\cite{Roper}
\ba
\int_k \psi_{Nj'}(\bar P_+,k)
\psi_{Nj}(\bar P_-,k)  =0,
\label{eqOrthCond2}
\ea
where we recall that $\bar P_\pm$
are the momenta for the case $Q^2=0$.\footnote{Considering
for instance the rest frame of the final
state for \mbox{$Q^2=0$,} one has
\ba
\bar P_+ = (M_{Nj'},0,0,0), \hspace{.2cm}
\bar P_- = \left(\frac{M_{Nj'}^2 + M_{Nj}^2}{2 M_{Nj'}},0,0,
-\frac{M_{Nj'}^2 - M_{Nj}^2}{2 M_{Nj'}}\right),
\nonumber
\ea
defining the three-momentum along $z$.
The sign of the $z$-component is chosen in order
to obtain $q$ with a positive sign along with the $z$-axis.}

Since the zero overlap between the radial
wave functions is equivalent to the
orthogonality between the radial wave functions
due to Eq.~(\ref{eqOrthCond2}),
to discuss how the orthogonality is assured,
and how the parameters of the
wave functions $\psi_{N1}$ and $\psi_{N2}$
given by Eqs.~(\ref{eqPsiN1}) and (\ref{eqPsiN2})
are determined,
we define the integral function:
\ba
{\cal I}_{j' j} (Q^2)
= \int_k \psi_{Nj'} (P_+,k) \psi_{Nj} (P_-,k).
\label{eqIj'j}
\ea
Note that, with the above notation
one has from Eq.~(\ref{eqI20}):
${\cal I}(Q^2) \equiv {\cal I}_{20}(Q^2)$.
As for ${\cal I}$, the integrals ${\cal I}_{j' j} $
are covariant and frame independent.
In the following we will consider only the case $Q^2=0$.

The orthogonality between $N1$ and $N0$ can be
imposed by the condition ${\cal I}_{10}(0)=0$, which
can be used to calculate the value of $\beta_3$.
The result obtained by this condition is $\beta_3= 0.1300$~\cite{Roper}.

We discuss now the orthogonality between
the $N2$ and the other states, $N1$ and $N0$.
Using the notation of Eq.~(\ref{eqIj'j}),
we can write the orthogonality with $N0$ and $N1$ by
the two conditions:
\ba
{\cal I}_{20}(0)=0,
\hspace{1cm}
{\cal I}_{21}(0)=0.
\label{eqN2orth}
\ea
Choosing proper frames for each integral,
we can reduce Eqs.~(\ref{eqN2orth}) to a system of two
equations with two unknowns, $\beta_4$ and $ \beta_5$.
The results obtained by solving the set of the two equations are
$\beta_4 = 0.3377$ and $\beta_5= 0.00855$.

\section{Results}
\label{secResults}

\begin{figure}[t]
\vspace{.3cm}
\centerline{
\mbox{
\includegraphics[width=2.8in]{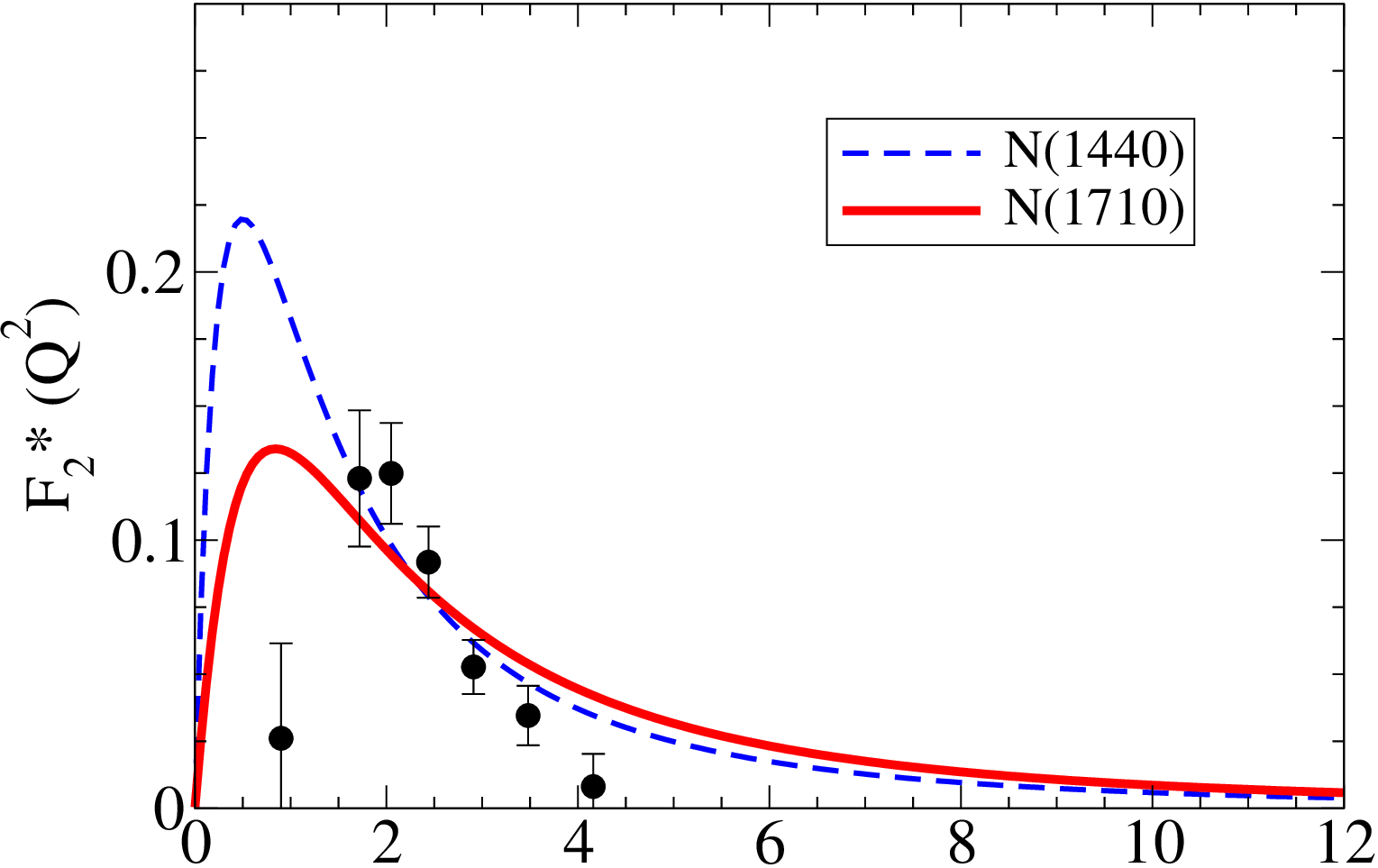}
}}
\centerline{
\vspace{.5cm} }
\centerline{
\mbox{
\includegraphics[width=2.8in]{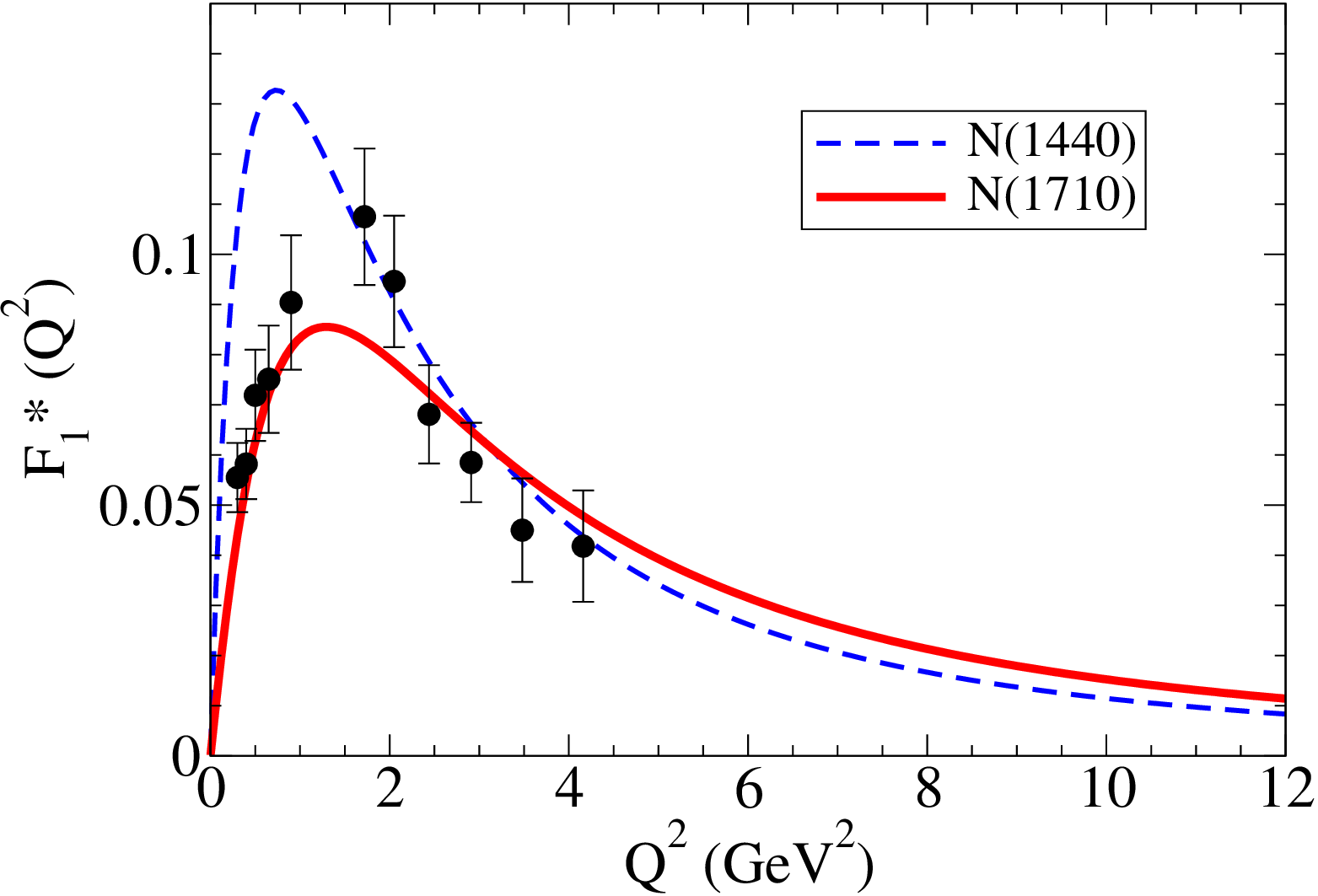}
}}
\caption{\footnotesize{
$\gamma^\ast N \to N(1710)$ transition form factors
(solid line) compared with
those for the Roper, $\gamma^\ast N \to N(1440)$
(dashed line).
Data for the Roper are from (CLAS)~\cite{CLAS}.}}
\label{figFF1}
\end{figure}

\begin{figure}[t]
\vspace{.3cm}
\centerline{
\mbox{
\includegraphics[width=2.8in]{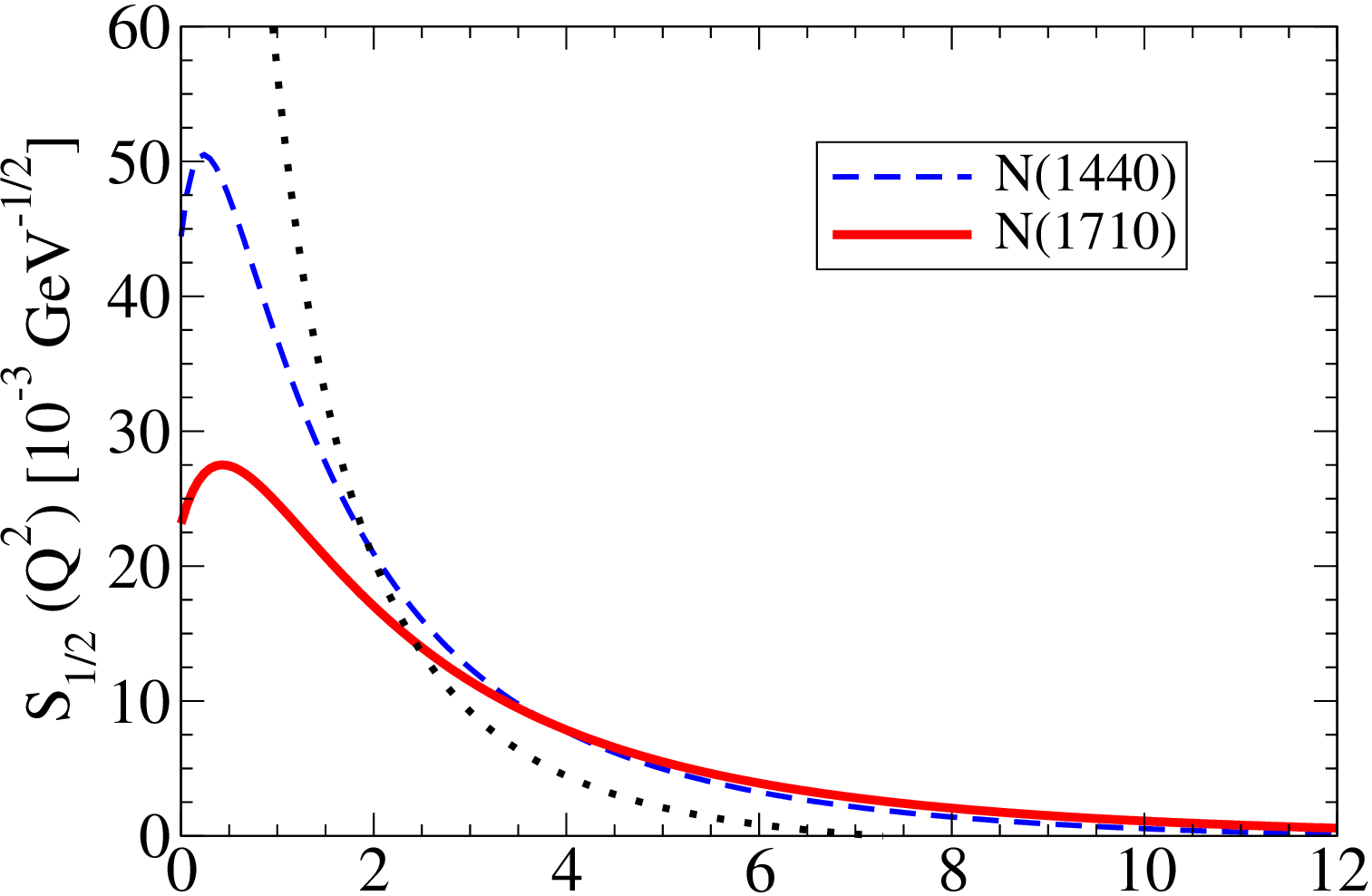}
}}
\centerline{
\vspace{.5cm} }
\centerline{
\mbox{
\includegraphics[width=2.8in]{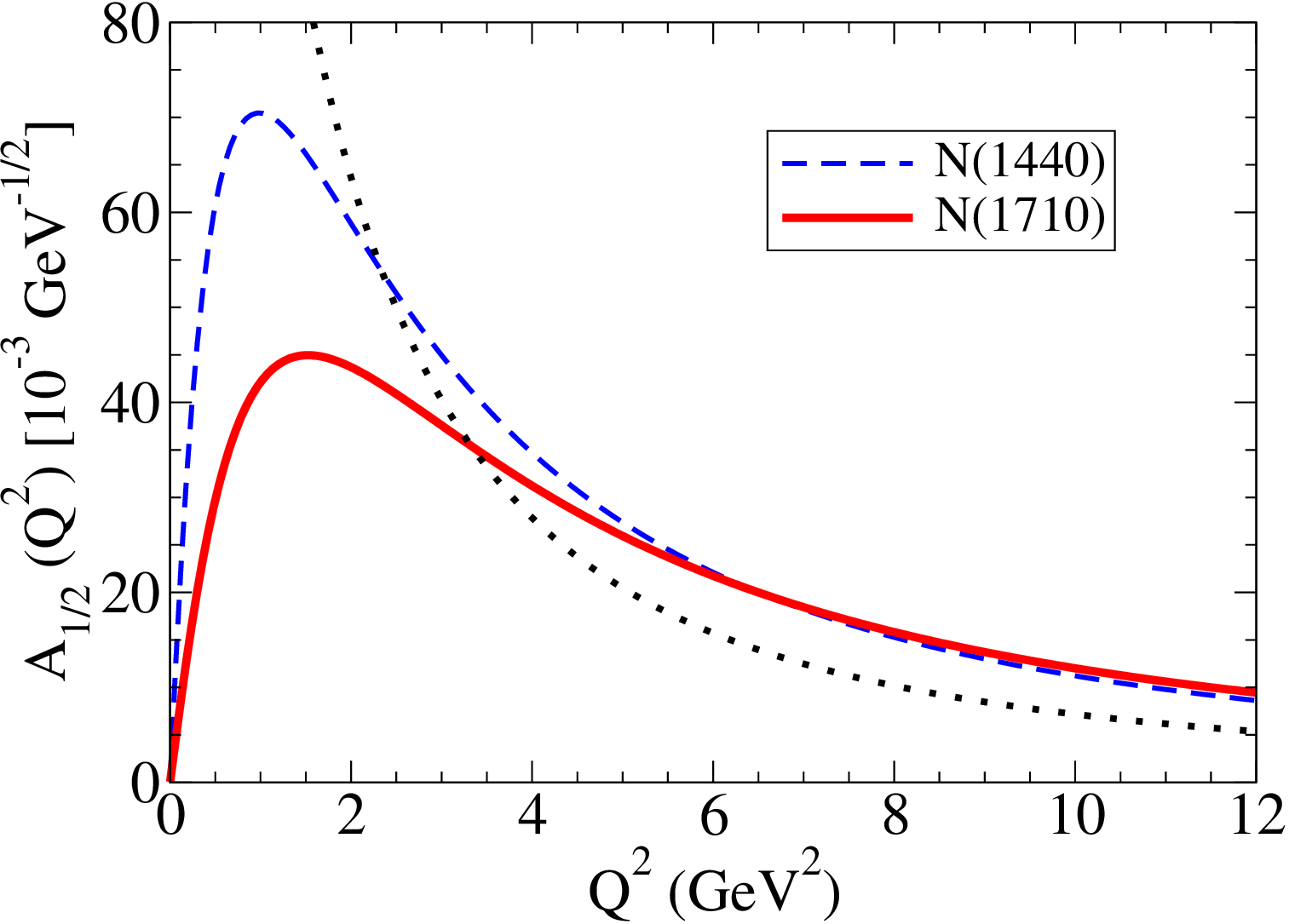}
}}
\caption{\footnotesize{
$\gamma^\ast N \to N(1710)$ helicity amplitudes
at the resonance rest frame (solid line).
Results are compared with those for the $\gamma^\ast N \to N (1440)$
amplitudes (dashed line),
and the {\it equivalent} amplitude for the nucleon $N(939)$ (dotted line).
See the text for the explanation on the {\it equivalent} nucleon case.
}}
\label{figAmp1}
\end{figure}

In the present work we consider only the reaction
with the proton ($N=p$), since
there are no data for the neutron
($N=n$) for finite $Q^2$, and
our model is expected to work better
in the region of large $Q^2$.
The results for the form factors are
presented in Fig.~\ref{figFF1}, and
those for the helicity amplitudes
are presented in Fig.~\ref{figAmp1},
both up to $Q^2=12$ GeV$^2$.

In Fig.~\ref{figFF1}, we also include
the results for $\gamma^\ast N \to N(1440)$,
obtained in Ref.~\cite{Roper}.
From Fig.~\ref{figFF1}, it is clear that
both reactions yield very close results
for the form factors $F_1^*$ and $F_2^*$,
in the region $Q^2 > 2$ GeV$^2$.
Recall that it is in the high $Q^2$
region that our model is more reliable,
since the valence quark degrees of freedom
are expected to be dominant.
In Fig.~\ref{figFF1} we also include the experimental data
for the $\gamma^\ast N \to N(1440)$ reaction from
CLAS for comparison (dataset with the highest value for $Q^2$).
It is appreciable from Fig.~\ref{figFF1}
that the agreement of the model (dash line)
and the data is very good for the higher $Q^2$ region ($Q^2 > 2$ GeV$^2$).
Only for the last $Q^2$ datapoint 
we can observe for $F_2^\ast$ a larger difference, 
about 2 standard deviations,
between the model result and the data.
New data for the higher $Q^2$ region are necessary
to test the model in more detail.
Other datasets are not included
since they are restricted to the low $Q^2$ region,
the region dominated by the meson cloud effects,
and therefore we should expect deviations
from our results~\cite{Aznauryan12a,Roper,RoperProc}.

The results for the amplitudes in Fig.~\ref{figAmp1} are calculated
using the form factors given in Eqs.~(\ref{eqA12}) and (\ref{eqS12})
for both cases, $N(1440)$ and $N(1710)$.
The PDG result for the $N(1710)$ case,
$A_{1/2}(0)= (24\pm 10)\times 10^{-3}$ GeV$^{-1/2}$~\cite{PDG},
is not included, since we are focusing on
the larger $Q^2$ region.
Note that the amplitudes
for the two reactions are even closer
than the results for the form factors,
particularly for $Q^2 > 4$ GeV$^2$.

It is possible that the closeness of the results
for these reactions in the higher $Q^2$ region
is a consequence
of the forms of the
radial wave functions (\ref{eqPsiN0}), (\ref{eqPsiN1})
and (\ref{eqPsiN2}), where  the radial wave functions of the
excited states are related with
the ground state nucleon radial wave function.
Note that all the radial wave functions
are parameterized with the same short range structure 
given by the factor $1/(\beta_2 + \chi)$,
apart from the mass differences in $\chi$.
Since the coupling with photon with high $Q^2$
probes the short range structure of the baryon
states it is expected that the form factors at high $Q^2$
have the same shape since they are described by 
the same parametrization.

In order to examine the above arguments in more detail, 
we plot also the {\it equivalent} results for the nucleon case (doted line).
Note that, according to Eqs.~(\ref{eqA12}) and (\ref{eqS12})
the amplitudes for the nucleon are
proportional to $G_M$ ($A_{1/2}$)
and $G_E$ ($S_{1/2}$),
but also there is an extra factor ${\cal R}$
depending on $K= \frac{M_R^2-M^2}{2 M_R}$,
that vanishes and induces singularities in the amplitudes
in the elastic limit ($M_R=M$).
In order to be able to compare the results of
the nucleon with those of the $N(1440)$ and $N(1710)$,
we keep the expression of ${\cal R}$
given for the $N(1440)$ and
take the $M_R \to M$ limit
in the factor  $|{\bf q}| \frac{M_R + M}{Q^2}$,
giving $\sqrt{\frac{1 + \tau}{\tau}}$,
in the case of $S_{1/2}$.
From Fig.~\ref{figAmp1} we can see
that, apart from the magnitude (about $1/\sqrt{2}$ smaller),
the falloff of the nucleon {\it equivalent} amplitudes
are about the same as those of the $N(1440)$ and $N(1710)$.

\subsection{High $Q^2$ parametrization}

To make a comparison easier with the expected
future experimental data, we parameterize our model
results using the simple analytic form,
\ba
& &
A_{1/2} = A \left(
\frac{\Lambda_1^2}{ \Lambda_1^2 + Q^2} \right)^{3/2},
\label{eqA120}
\\
& &
S_{1/2} = S \left(
\frac{\Lambda_2^2}{ \Lambda_2^2 + Q^2} \right)^{5/2},
\label{eqS120}
\ea
which is consistent with the falloff expected
for very large $Q^2$~\cite{Carlson}.
In Eqs.~(\ref{eqA120}) and~(\ref{eqS120}), $A$ and $S$
are constants, and $\Lambda_1^2$ and $\Lambda_2^2$ are cutoffs squared.

The numerical values for the parametrization
are given in Table~\ref{tableLargeQ2}.
The parameters are determined so as to
reproduce the results exactly at $Q^2 = 6$ GeV$^2$,
but provide also good approximations for the values
of $Q^2$ up to 10 GeV$^2$.
The exception is the parametrization for the
amplitude $S_{1/2}$ in the $\gamma^\ast N \to N(1440)$
reaction, that approaches zero very fast.
In this case the approximation is valid only around $Q^2=5$ GeV$^2$.
Note that the expressions (\ref{eqA120}) and (\ref{eqS120})
are only valid in the range of $Q^2$
presented in the figures.
In general the covariant spectator quark model
has smooth logarithmic corrections
for the form factors and helicity amplitudes~\cite{Nucleon,Roper}.
Therefore the parametrizations given by Eqs.~(\ref{eqA120})-(\ref{eqS120})
and Table~\ref{tableLargeQ2}, are not valid
for an arbitrary high $Q^2$.

Since our results for the form factors
are well approximated by the
parametrizations~(\ref{eqA120})-(\ref{eqS120})
for the region 5--10 GeV$^2$,
we can expect evidences of asymptotic behavior
for $Q^2 >$ 5 GeV$^2$.

\subsection{Discussion of results}

The $\gamma^\ast N \to N(1710)$ reaction
was studied in the past within several frameworks.
We start discussing the results from
quark models, and later discuss other models.

Analytic expressions for the
$\gamma^\ast N \to N(1710)$ helicity amplitudes
derived from an algebraic model of QCD,
also based on the quark degrees of freedom,
can be found in Ref.~\cite{Bijker}.
Estimates from quark models
are given in Refs.~\cite{Capstick95,Santopinto12,Ronniger13}.
The magnitudes of those estimates are consistent
with our predictions.
In general the amplitudes $A_{1/2}$ and $S_{1/2}$ are positive.
The exception is Ref.~\cite{Santopinto12},
where $S_{1/2}$ is negative.

Some others models describe the $N(1710)$ state
as a quark bare core with some excitations~\cite{Golli08,Alberto01}.
In Ref.~\cite{Golli08}
a coupled-channel formalism is
combined with a description of
the baryon bare cores and meson production
based on the cloudy bag model.
In Ref.~\cite{Alberto01}
the $N(1710)$ state is described by a baryon bare core
with some radial excitations, combined
with glueball and $\sigma$ vibrational excitations.
In both cases~\cite{Golli08,Alberto01}, we can expect a dominance
of the bare core in the high $Q^2$ region, since
meson cloud is expected to be suppressed.

In other frameworks the $N(1710)$ state is
dynamically generated
from some baryon and meson
states~\cite{Khemchandani08,EBAC1,Suzuki10,Zhang12}.

The EBAC model~\cite{EBAC1,Suzuki10} uses baryon-meson coupled-channel
formalism to describe the photo- and electro-production
of mesons by nucleons.
In that framework the $N(1710)$,
identified as $N(1820)$, is a state that
evolves from an $N(1763)$ bare core state through
its coupling with the $\pi N, \eta N$ and $\pi\pi N$ channels.
The EBAC model is also used to extract
the amplitude $A_{1/2}$ from the data,
for $Q^2 < 1.5$ GeV$^2$ in Ref.~\cite{Suzuki10}.
Although their results cannot be compared
with our results which are reliable only for the
high $Q^2$ region, we note that the real part of
the amplitude they extracted is positive,
with the magnitude comparable with ours.

In Ref.~\cite{Khemchandani08} by solving the Faddeev equations
with the input of the two-body coupled-channel t-matrices,
the $N(1710)$ state is explained as a
dynamical generated resonance dominated
by the $\pi \pi N$ component (the $\pi \pi$ component
can be also interpreted as a $\sigma$-meson).
Also in Ref.~\cite{Zhang12} the $N(1710)$
emerges as the result of the pion dressing
of nucleon and $\Delta$ bare cores.
The model parameters were fixed to reproduce
masses and branching ratios.

In models that the $N(1710)$ state is generated
by the baryon-meson interactions
as the ones mentioned above~\cite{Khemchandani08,EBAC1,Suzuki10,Zhang12},
we can expect a much larger spatial extended structure for $N(1710)$,
and consequently a faster falloff
with $Q^2$ for the associated transition
amplitudes~\cite{Bijker2}.

From the discussions above,
we conclude that we can deduce the
nature of the $N(1710)$ state by
the $Q^2$ behavior of the helicity amplitudes,
which may be revealed in future experiments.
Assuming a dominance of the valence quark effects as we do,
we  expect $A_{1/2} \propto 1/Q^3$
and $S_{1/2} \propto 1/Q^5$ in the very large $Q^2$ region.
Instead, if the $N(1710)$ system is dominated
by $qqq$-$(q\bar q)$ configurations,
the falloffs mentioned previously are modified by an extra
$1/Q^4$ factor according to pQCD~\cite{Carlson}.

A few words are in order on the possible
existence of another $N\frac{1}{2}^+$ state
with a invariant mass near 1.7 GeV.
PDG reports a forth $N\frac{1}{2}^+$ state
(two-star resonance) denoted by $N(1880)$.
However, several partial wave analyses and
coupled-channel reaction models
suggest the existence
of another $N\frac{1}{2}^+$ state near the mass 1.7 GeV.
See for instance
Refs.~\cite{Ceci06a,Ceci06b,Osmanovic11,Ronchen13,Batinic95,Chiang04}.
If there exist two $N\frac{1}{2}^+$ states with very close
masses around 1.7 GeV,
the experimental determination of the individual contributions
will be very difficult.
Then, the transition helicity amplitudes
extracted from the data would be
a combination of the two resonances,
and our model, that assumes only
one radial excited state, would fail
the description of the data.

\begin{table*}[t]
\begin{tabular}{c c c c c}
\hline
\hline
  & $A (10^{-3} \mbox{GeV}^{-1/2})$ & $\Lambda_1^2 (\mbox{GeV}^2)$
  & $S (10^{-3} \mbox{GeV}^{-1/2})$ & $\Lambda_2^2 (\mbox{GeV}^2)$ \\
\hline
$N(1440)$ & 275.40 &  1.374 &  {\it 727.53} & {\it 0.789} \\
$N(1710)$ & 122.17 & 2.783 &  210.71 & 1.531 \\
\hline
\hline
\end{tabular}
\caption{
Parametrization for the helicity amplitudes for the higher $Q^2$ region
according to Eqs.~(\ref{eqA120}) and (\ref{eqS120}).
See discussion in the main text.}
\label{tableLargeQ2}
\end{table*}

\section{Summary and conclusions}
\label{secConclusions}

The future JLab-12 GeV facility will open a possibility
of studying in detail the electromagnetic
structure of the resonances in the third resonance region.
There is then the chance of
exploring the resonances
$N(1720)\frac{3}{2}^+$,
$\Delta(1720)\frac{3}{2}^-$  (four-star resonances),
$N(1710)\frac{1}{2}^+$, and
$N(1700)\frac{3}{2}^-$
(three-star resonances).
Among these the $N(1710)\frac{1}{2}^+$
is a very interesting system
due to the significant branching ratios to
the $\eta N, K\Lambda$ and $\pi \pi N$ channels.
Some authors defend even that the state should
be reclassified as a four-star resonance~\cite{Ceci06a}.

Future data for the invariant mass $W \approx 1.7$ GeV
will help to clarify the nature of the state $N(1710)\frac{1}{2}^+$,
namely, if it is a radial excitation of the nucleon and the Roper,
or a more complex and/or exotic system.
From the observation
of the measured helicity amplitudes
in the large $Q^2$ region, we will be able to draw
some conclusion on the dominant degrees of freedom
in the system at large $Q^2$.
If $N(1710)$ is dominated by the valence quark
degrees of freedom only, we should observe
the scaling behavior with $Q^2$ as $A_{1/2} \propto 1/Q^3$.
In the case of baryon-meson {\it molecular} system
we expect a falloff of  $A_{1/2}$ faster than $1/Q^3$.
In addition,
in principle it will also be possible
to answer the question if
there is the another $N\frac{1}{2}^+$ state
close to the one that we study in this work.

In this work we have used the covariant spectator quark model,
to predict the transition form factors and
the helicity amplitudes
for the $\gamma^\ast N \to N(1710)$ reaction.
The covariant spectator quark model
was already applied to several nucleon resonances successfully.
Since we have not included the meson cloud effects
which are known to be very important in the low $Q^2$ region,
we expect our prediction to the
$\gamma^\ast N \to N(1710)$ helicity amplitudes to be valid
only in the high $Q^2$ region,
where the valence quark degrees of freedom are dominant.

We have assumed that the $N(1710)$ state is the second radial
excitation of the nucleon, similar to the assumption that
the Roper is the first radial excitation of the nucleon.
The wave function of the $N(1710)$
is uniquely determined, apart from the sign,
by the orthogonality of the $N(1710)$ wave function
with the nucleon and $N(1440)$ wave functions.
The nucleon and $N(1440)$ wave functions were
determined in the previous works~\cite{Nucleon,Roper}.

For high $Q^2$, particularly for
$Q^2 > 4$ GeV$^2$, the calculated helicity amplitudes
for the $\gamma^\ast N \to N(1710)$ reaction
show results that are very close to those for
the $\gamma^\ast N \to N(1440)$ reaction.
Therefore, the future measurements
of the helicity amplitudes in the large $Q^2$ region
can be used to test the assumption that the
$N(1710)$ state is the second radial excitation of the nucleon.

Finally we note that the present formalism
can be used to study the radial excited states
in the $\Delta$ sector (isospin 3/2)~\cite{Delta1600},
and also in the strange baryon sector.
In the latter case we can study the transitions
$\gamma^\ast \Lambda \to \Lambda (1600) \frac{1}{2}^+$,
$\gamma^\ast \Lambda \to \Lambda (1810) \frac{1}{2}^+$ and
$\gamma^\ast \Sigma \to \Sigma (1660) \frac{1}{2}^+$
(all three-star resonances).
More experimental support is necessary also in these cases.

\begin{acknowledgments}
The authors would like to thank
Ralf Gothe and Catarina Quintans
for helpful discussions.
This work was supported by the Brazilian Ministry of Science,
Technology and Innovation (MCTI-Brazil), and
Conselho Nacional de Desenvolvimento Cient{\'i}fico e Tecnol\'ogico
(CNPq), project 550026/2011-8.
\end{acknowledgments}

\appendix

\section{Covariant spectator quark model}
\label{appSQM}

In the following we present some details of
the covariant spectator quark model.

\subsection{Quark form factors}

\begin{table}[t]
\begin{center}
\begin{tabular}{c c c c c c c}
\hline
\hline
$\kappa_+$ & $\kappa_-$ & $c_+$ & $c_-$ & $d_+$ & $d_-$ & $\lambda_q$ \\
\hline
1.639 & 1.823 & 4.16 & 1.16 & $-0.686$ & $-0.686$ & 1.21 \\
\hline
\hline
\end{tabular}
\end{center}
\caption{Parameters in the quark current.}
\label{tabParam}
\end{table}

The quark current associated with Eq.~(\ref{eqJi}) is expressed
in terms of the quark form factors $f_{i\pm}$ ($i=1,2$), inspired by
a vector meson dominance form:
\ba
& &
\hspace{-1.2cm}
f_{1\pm}(Q^2)=
\lambda_q + (1-\lambda_q)\frac{m_v^2}{m_v^2+Q^2}
+ c_\pm \frac{M_h^2Q^2}{(M_h^2+Q^2)^2}, \\
& &
\hspace{-1.2cm}
f_{2\pm} (Q^2)=
\kappa_\pm
\left\{
d_\pm
\frac{m_v^2}{m_v^2+Q^2}+
(1-d_\pm)\frac{M_h^2}{M_h^2+Q^2}
\right\}.
\ea
In the above, $\lambda_q$ defines
the quark charge in deep inelastic scattering,
$\kappa_\pm$ are the isoscalar and isovector
quark anomalous magnetic moments.
The mass $m_v$ ($M_h$) corresponds to
the light (heavy) vector meson,
and $c_\pm$, $d_\pm$ are the mixture coefficients.
In the present model we set $m_v =m_\rho$ ($\approx m_\omega$)
for the light vectorial meson
and $M_h = 2M$ (twice the nucleon mass)
to represent the short-range physics.
The values of the parameters were previously fixed by
the nucleon elastic form factors~\cite{Nucleon},
and they are presented in
Table~\ref{tabParam}. Note that the present model uses $d_+=d_-$.

\subsection{Wave functions}

In the covariant spectator quark model the $S$-state wave functions
for the nucleon and the
nucleon radial excitations $\Psi_{Nj}(P,k)$
can be represented by Eq.~(\ref{eqPsiNj})
for the states labeled by $Nj$
with $j=0,1,2$~\cite{Nucleon,Roper,OctetFF}.
In Eq.~(\ref{eqPsiNj}) the isospin operators $\phi_I^{0,1}$
act on the $Nj$'s isospin states $\chi^t$, where
$\chi^{+1/2}= (\begin{array}{cc} 1 &0  \cr \end{array} )^T$
and $\chi^{-1/2}= (\begin{array}{cc} 0&1 \end{array})^T$.
The explicit form are~\cite{Nucleon,Nucleon2}
\ba
\phi_I^0 \chi^t = \unit \chi^t, \hspace{1cm}
\left(\phi_I^1\right)_l \chi^t =
- \frac{1}{\sqrt{3}} (\tau \cdot \xi_l^\ast) \chi^t,
\ea
where $\xi_l$ is the isospin vector of the isospin-one diquark, defined
in a usual way (in the spherical basis),
\ba
\xi_\pm =
\mp \frac{1}{\sqrt{2}}
\left(
\begin{array}{c} 1 \cr \pm i \cr 0 \end{array} \right),
 \hspace{1cm}
\xi_0 =
\left(
\begin{array}{c} 0 \cr 0 \cr 1 \end{array} \right).
\ea

As for the spin states $\phi_S^{0,1}$, we start
to present the spin-1 polarization vector $\varepsilon_{\Lambda P}$
in a Fixed-Axis base for the case of a baryon with momentum
$P=\left(E_B,0,0,P_z \right)$,
where $E_B=  \sqrt{M_B^2 + P_z^2}$ ~\cite{FixedAxis,Nucleon2}:
\ba
& &
\varepsilon_{\pm P}^\alpha =
\mp
\frac{1}{\sqrt{2}}
(0,1,\pm i,0), \nonumber \\
& &
\varepsilon_{0 P}^\alpha =
\left( \frac{P_z}{M_B},0,0, \frac{E_B}{M_B}\right).
\ea
We can then write
\ba
& &
\phi_I^0 = u_B(P), \nonumber \\
& &
\phi_I^1 = - (\varepsilon_{\Lambda P}^\ast)_\alpha U_B^\alpha(P),
\ea
where $u_B$ is a Dirac spinor, and
\ba
U_B^\alpha (P)=
\frac{1}{\sqrt{3}} \gamma_5 \left(
\gamma^\alpha - \frac{P^\alpha}{M_B}\right)u_B(P).
\ea
In the case of the nucleon one may replace,
$u_B \to u$ and $M_B \to M$.
In the case of a resonance $R$
one may replace the index $B$ by the index $R$.

\end{document}